\documentclass[a4paper,11pt]{article}
\pdfoutput=1 

\usepackage{jinstpub} 

\usepackage{lineno}
\usepackage{hyperref}
\usepackage{subcaption}

\title{Charge exchange radiation diagnostic with gas jet target for measurement of plasma flow velocity in the linear magnetic trap}

\author{A. Lizunov\note{Corresponding author.}}
\affiliation{Budker Institute of nuclear physics,\\630090 Novosibirsk, Russia}

\emailAdd{lizunov@inp.nsk.su}

\abstract{The ambipolar electrostatic potential rising along the magnetic field line from the grounded wall to the centre in the linear gas dynamic trap, rules the available suppression of axial heat and particle losses. In this paper, the visible range optical diagnostic is described using the Doppler shift of plasma emission lines for measurements of this accelerating potential drop. We used the room temperature hydrogen jet puffed directly on the line of sight as the charge exchange target for plasma ions moving in the expanding flux from the mirror towards the wall. Both bulk plasma protons and $He^{2+}$ ions velocity distribution functions can be spectroscopically studied; the latter population is produced via the neutral He tracer puff into the central cell plasma. This way, potential in the centre and in the mirror area can be measured simultaneously along with the ion temperature. A reasonable accuracy of $4\div8\%$ was achieved in observations with the frame rate of $\approx 1~kHz$. Active acquisitions on the gas jet also provide the spatial resolution better than 5~mm in the middle plane radial coordinate because of the strong compression of the object size when projected to the centre along the magnetic flux surface. The charge exchange radiation diagnostic operates with three emission lines: H-$\alpha$ 656.3~nm, He-I 667.8~nm and He-I 587.6~nm. Recorded spectra are shown in the paper and examples for physical dependences are presented. The considered experimental technique can be scaled to the upgraded multi-point diagnostic for the next generation linear traps and other magnetic confinement systems.}


\arxivnumber{1234.56789} 

\begin{document}
\maketitle
\flushbottom

\section{Introduction}
\label{sec:intro}
Linear magnetic systems for plasma confinement, also frequently referred as open-ended traps, have the common issue of field lines facing the grounded metallic wall somewhere beyond the mirror. The particular magnetic field configuration for different devices varies. In order for these confinement concepts to be attractive for real applications, the axial heat flux through the direct contact with the wall must be radically depressed comparing to the classic Spitzer \cite{spitzer} heat conductivity. The gas dynamic trap (GDT) \cite{gdt-review-ppcf} utilizes a strongly expanding magnetic "fan" beyond the mirror with a straight or curved inwards field line shape. The axial profile of the plasma electrostatic potential plays a crucial role forming the actual heat transport physics. In a steady state, this ambipolar potential equalises the electron and ion currents onto the wall. Study of the axial particle and energy transport~\cite{nf-axconf-2020} is one of the top priorities in the GDT scientific task list. This activity embraces new diagnostics development as well as experimental and theoretical research. 

\begin{figure}[htbp]
\centering 
\includegraphics[width=.9\textwidth]{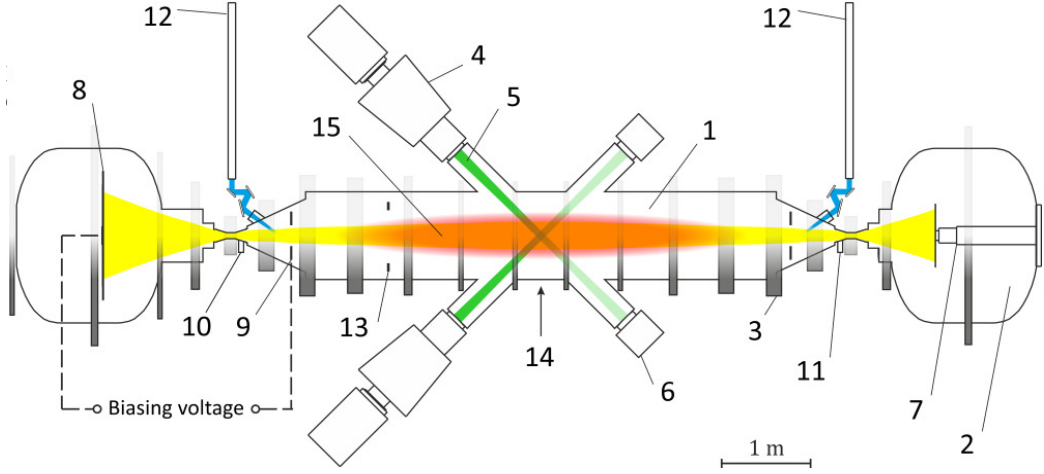}
\caption{\label{fig:gdt} The gas dynamic trap: 1 -- central cell, 2 -- right expander tank, 3 -- magnetic coil of central solenoid, 4 -- atomic beam injector, 5 -- deuterium beam, 6 -- beam dump, 7 -- arc discharge plasma source, 8 -- plasma dump in the left expander tank, 9 -- radial limiter, 10 -- left gas box, 11 -- right gas box,  12 -- waveguides of ECRH system, 13 -- diamagnetic loop, 14 -- Thomson scattering diagnostic.}
\end{figure}

Layout of the GDT device in the Budker Institute is shown in the Figure~\ref{fig:gdt}. The detailed description of plasma heating and sustainment scenarios can be found in \cite{gdt-review-ppcf}, \cite{ecrh_prl}. 
The current GDT scenario employs electron cyclotron resonance (ECR) discharge for the plasma startup, but does not involve ECR heating. Only one of gyrotrons and waveguides (12) drawn in the Figure~\ref{fig:gdt}, is used. In this scenario, the energetic ion population and bulk plasma heating is produced via the neutral beam injection.

\section{Diagnostic setup}
\label{diagnostic}

The analysis of the velocity distribution function for ions streaming out of the magnetic mirror, would bring the desired information about the electrostatic potential drop along the trajectory and the ion temperature. In GDT plasmas, the particle flow through the mirror towards the absorber surface, is effectively collisionless. The ion energy and magnetic moment conservation in a collisionless regime is expressed as
\begin{equation}
\label{ion_energy}
\frac{m_i v^2}{2} = \frac{m_i v_0^2}{2} + \Delta U(z),
\qquad
 v_{\perp}^2 = v_{\perp_0}^2 \frac{H(z)}{H_0} \thickapprox 0.
\end{equation}
Here in \eqref{ion_energy}, $v$ and $v_0$ is the ion velocity in the measurement point in the expander and in the central cell, $U(z) = q\phi(z)$ is the potential energy for the $q$-charged ion in the electrostatic potential distribution $\phi(z)$. 
The $z=0$ point is the device mid-plane. We imply that the measurement location is far beyond the mirror, so $H(z) / H_0 \ll 1$ and one can assume $v \approxeq v_z$. 
The one half of the Maxwellian ion distribution function (IDF) with $v_z \geq 0$ leaves the trap through the mirror, accordingly IDF in some point within the expansion region is
\begin{equation}
\label{idf}
f_i = n \left( \frac{m_i}{2\pi T_i} \right)^{3/2}e^{U/T_i}\exp \left(-\frac{m_i v^2}{2T_i} \right).
\end{equation}
Acceleration in the potential drop means that \eqref{idf} is not zero only for the axial velocity above the value defined by the expression
\begin{equation}
\label{vz_min}
\frac{m_i v_{z_{min}}^2}{2} = q\Delta\phi.
\end{equation}
The equation \eqref{vz_min} already provides the measurement approach. A natural and commonly used way to measure the axial ion velocity (or energy) in the plasma flux is by means of a grid electrostatic analyser, where the scanning analysing voltage is applied to decelerate ions. In past experiments in GDT, such experiments were successfully done~\cite{gdt-review-ppcf} (page~26). There are some flaws though in this technique. For example, it is typically tricky to arrange measurements in multiple radial points with the grid energy analyser. A special high voltage power supply for such an analyser can be complex and expensive. In this paper, we are considering a spectroscopic approach to the task. The method relies on charge exchange conversion of streaming ions into atoms with the subsequent light emission, which implies the classical Charge eXchange Radiation Spectroscopy (CXRS) scheme:
\begin{equation}
\label{cxrs}
A^{Z+} + H^0 \rightarrow A^{*(Z-1)+} + H^+ \rightarrow A^{(Z-1)+} + H^+ + h\nu.
\end{equation}
The neutral hydrogen target is used in \eqref{cxrs}, which is typical for CXRS. The emitted light spectrum shape encodes the IDF parameters. The ion temperature can be calculated by the Doppler FWHM (Full Width Half Maximum) as
\begin{subequations}
\label{cxrs_doppler}
\begin{align}
\label{cxrs_doppler_width}
T_i = \frac{m_i c^2}{2\ln2}\left( \frac{\delta\lambda_{1/2}}{2\lambda_0} \right)^2,
\end{align}
where $\lambda_0$ is the unshifted wavelength. In turn, the accelerating potential drop is linked to the Doppler line shift as
\begin{equation}
\label{cxrs_doppler_shift}
\Delta\phi = \frac{m_i c^2}{2q}\left( \frac{\Delta\lambda_D}{\lambda_0} \right)^2 \frac{1}{\cos^2\Theta},
\end{equation}
\end{subequations}
where $\Theta$ is the angle between the ion velocity and the Line of Sight (LOS) direction.

\begin{figure}[htbp]
\centering 
\includegraphics[width=.7\textwidth]{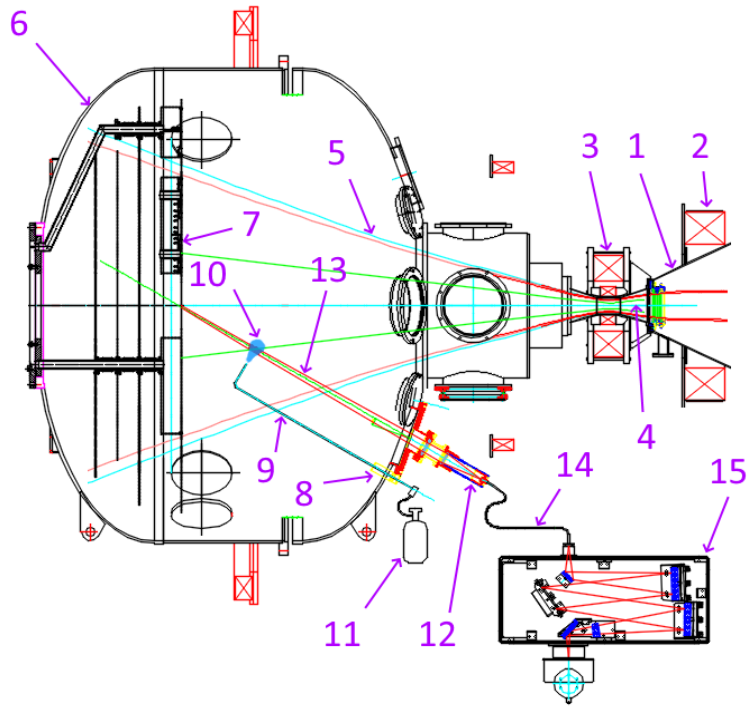}
\caption{\label{fig:w-exp_elvis} Layout of CXRS measurements in the left GDT expander: 1 -- cone part of the GDT central cell, 2 -- magnetic coil, 3 -- mirror magnetic coil assembly, 4 -- gas puff volume, 5 -- boundary magnetic field line, 6 -- expander tank, 7 -- plasma dump, 8 -- translation vacuum feedthrough of the gas feed tube, 9 -- quartz tube 2~mm diameter, 10 -- charge exchange gas target, 11 -- $H_2$ reservoir and the feed line with the pulsed electromagnetic valve, 12 -- 2-inch lens for light collection, 13 -- light collection solid angle, 14 -- optical fibre optic, 15 -- spectrometer coupled with the CCD camera.}
\end{figure}

Diagnostics for CXRS in a tokamak or other magnetic plasma confinement device, are typically associated with the atomic (hydrogen or deuterium) beam acting as a target. Indeed it is generally a crucial  requirement to deliver a substantial target atom density to the certain point inside the bulk of the plasma. The major performance criterion here is the signal-above-background ratio (S/B). In the simplest case, it is given by the relation between the active CX signal recorded on the target, and the passive emission collected along the LOS. A respectable $S/B\gtrsim 1$ can be achieved with a high enough beam current density in the measurement point and a sufficiently high particle energy to have a possibly low beam trapping on the way to this point. In some particular cases, supersonic gas jets \cite{sgi-1, sgi-2} or thermal gas puff can be used for optical diagnostics. A well established method for study of electron temperature and density in the scrape-off layer and edge turbulence in hot magnetically confined plasmas is based on relative intensities of neutral helium (He-I) lines \cite{he_ratio_nstx-u, he_ratio_rfx, he_ratio_aug, he_ratio_hl2a}. These diagnostics record light emitted by the introduced helium atoms. For our spectroscopy needs in GDT expander, the room temperature jet of molecular hydrogen is applied as the charge exchange target for plasma ions. The setup of CXRS measurements in the GDT expander is drawn in the Figure~\ref{fig:w-exp_elvis}.
The ion density (both for the plasma majority and impurities) in the expanding plasma flow drops with the expansion ratio $k = H(z)/H_0$ roughly linear. In the region under study, it is $50\div100$ less than that in the central cell, which poses a certain diagnostic challenge. On the other hand, restrictions against perturbing the plasma parameters are mild. It is proven \cite{gdt_expander_infulence} that even a strong emission of cold electrons downstream in the expander do not affect the central cell plasma. Then it is advantageous to use a gas jet puffing from the narrow capillary placed directly on the LOS instead of the energetic atomic beam because a much larger local atomic density (and the rate of CX events) can be achieved. The CXRS gas target assembly is a 2~mm-diameter quartz tube inserted via the vacuum feedthrough and connected to the pulsed fast electromagnetic valve. The valve open delay relative to the measurement time window, is adjusted to the minimum while it is still sufficient to generate an effective CX target. The main concern is to reduce the additional gas load in the expander tank during the plasma discharge and so to less contaminate other measurements in the expander. The estimated molecular hydrogen density of $n(H_2) \cong 10^{19}~m^{-3}$ is observed sufficient for an ample optical signal of CX emission, as it is shown below in the paper. The local observation volume (gas cloud) has the size of $\approx 20 mm$, but the actual space resolution is defined by the cloud size mapped onto the GDT central plane along the magnetic flux surface: $\delta r_0 \approx \delta r_{cloud}\sqrt{H(z) / H_0} \approx 2\div5 mm$. In the paper, radii are expressed in the device mid-plane if not otherwise specified. As the Figure~\ref{fig:w-exp_elvis} illustrates, the diagnostic LOS is fixed. The gas tube tip is movable between the axis and the plasma periphery along the LOS allowing for profile acquisition in a series of shots. One should keep in mind that both the radius and the z-coordinate are changed at the same time along the LOS. 

The optical system (12) (see Figure~\ref{fig:w-exp_elvis}) collects the light, which comes to the spectrometer (15) via the fibre optical light guide (14). The Table~\ref{tab:diag_param} summarizes the main parameters of the registration system. The spectrometer we used is the factory LOMO MDR-23 model with the custom cylindrical output lens for the astigmatism correction. It is coupled with the Princeton Instuments PyLoN CCD camera \cite{pylon-ccd} which has a small dark current and readout noise. The CCD is configured in the "Kinetics" mode (see \cite{pylon-ccd}) featuring multiple exposures on the sensor within the single digitization and the readout cycle. This regime provides a trade-off for a reasonably fast frame rate in the kHz region with a limited number of exposures at the price of sacrificing the light throughput. We set ten exposures of $0.5 ms$ duration and the effective frame rate of 1.1~kHz, the spectrometer entrance slit is accordingly enabled only on the $1/10$ of its height.
\begin{table}[htbp]
\centering
\caption{\label{tab:diag_param} Main parameters of the optical registration system.}
\smallskip
\begin{tabular}{| l | l |}
\hline
\emph{Spectrometer} & \\
\hline
Optical scheme & Czerny-Turner \\
Focal distance & 600 mm \\
F/No. & F/6 \\
Groove density & 1800 g/mm \\
Blaze & 550 nm \\
Dispersion & 0.69 nm/mm \\
Spectral resolution & 0.041 nm \\
\hline
\emph{CCD} & \\
\hline
Model & PyLoN 2KB eXcelon (LN-cooled to -120\textcelsius) \\
Sensor & 2048x512 pixels $13.5 \times 13.5 \mu m$ \\
Binning & \emph{Kinetics}, 50 rows vertical \\
Exposures on CCD & 10 \\
Exposure duration & 0.5 ms \\
Frame rate & 1.1 kHz \\
System noise & $\approx 3.5 e-$ \\
\hline
\end{tabular}
\end{table}

\section{Measurement of IDF by Doppler spectroscopy}
\label{results}
\paragraph{Hydrogen H-alpha spectral line.}
The observation geometry shown in the Figure~\ref{fig:w-exp_elvis} shows that the angle between the ion velocity vector and the LOS vector (the latter being directed towards the optics) $\Theta > 90^\circ$. This angle varies from $\Theta_0 = 150^\circ$ on the axis to the $\Theta_{edge} = 130^\circ$ on the edge field line projecting to $r_0 = 15 cm$ -- the radial limiter radius. For these angles, the Doppler shift is towards larger wavelengths. In all GDT experimental scenarios, there is a non-negligible hydrogen plasma component even if deuterium is used to create both the bulk plasma and fast ions via neutral beam injection. Possible explanations of that are some organic residuals inside the vacuum vessel and also micro-leaks of the air with water vapor. Anyway, observation of the Doppler-shifted D-$\alpha$ 656.1~nm spectral line is not expedient because the left wing of the cold-gas H-$\alpha$ 656.3~nm would overlap it. Instead, we modified the plasma scenario with the portion of hydrogen puffing from the left gas box, see (10) in the Figure~\ref{fig:gdt}. A typical molecule or atom free path before ionization amounts $\lesssim 10~mm$, so one can assume hydrogen ion birth places localized in the left mirror area. We will refer the plasma electrostatic potential calculated by the H-$\alpha$ Doppler shift as the "potential in the mirror" to distinguish from the central potential.
\begin{figure}[htbp]
\centering 
\includegraphics[width=.7\textwidth, trim=50 10 100 10,clip]{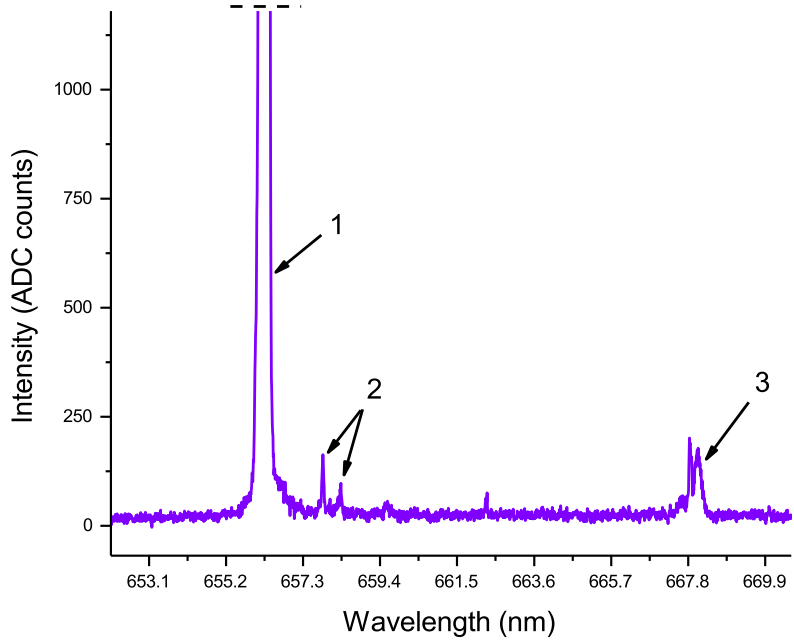}
\caption{\label{fig:full_spectrum} The spectrum recorded on the axis with the CX gas target in GDT shot 49311 at $t=8~ms$, exposure $\tau=0.5~ms$: 1 -- bright D-$\alpha$ and H-$\alpha$ lines (cut out), 2 -- C-II lines 657.8~nm and 658.3~nm, 3 -- He-I line 667.8~nm with Doppler-shifted component.}
\end{figure}
\begin{figure}[htbp]
\centering 
\includegraphics[width=.7\textwidth, trim=40 10 100 10,clip]{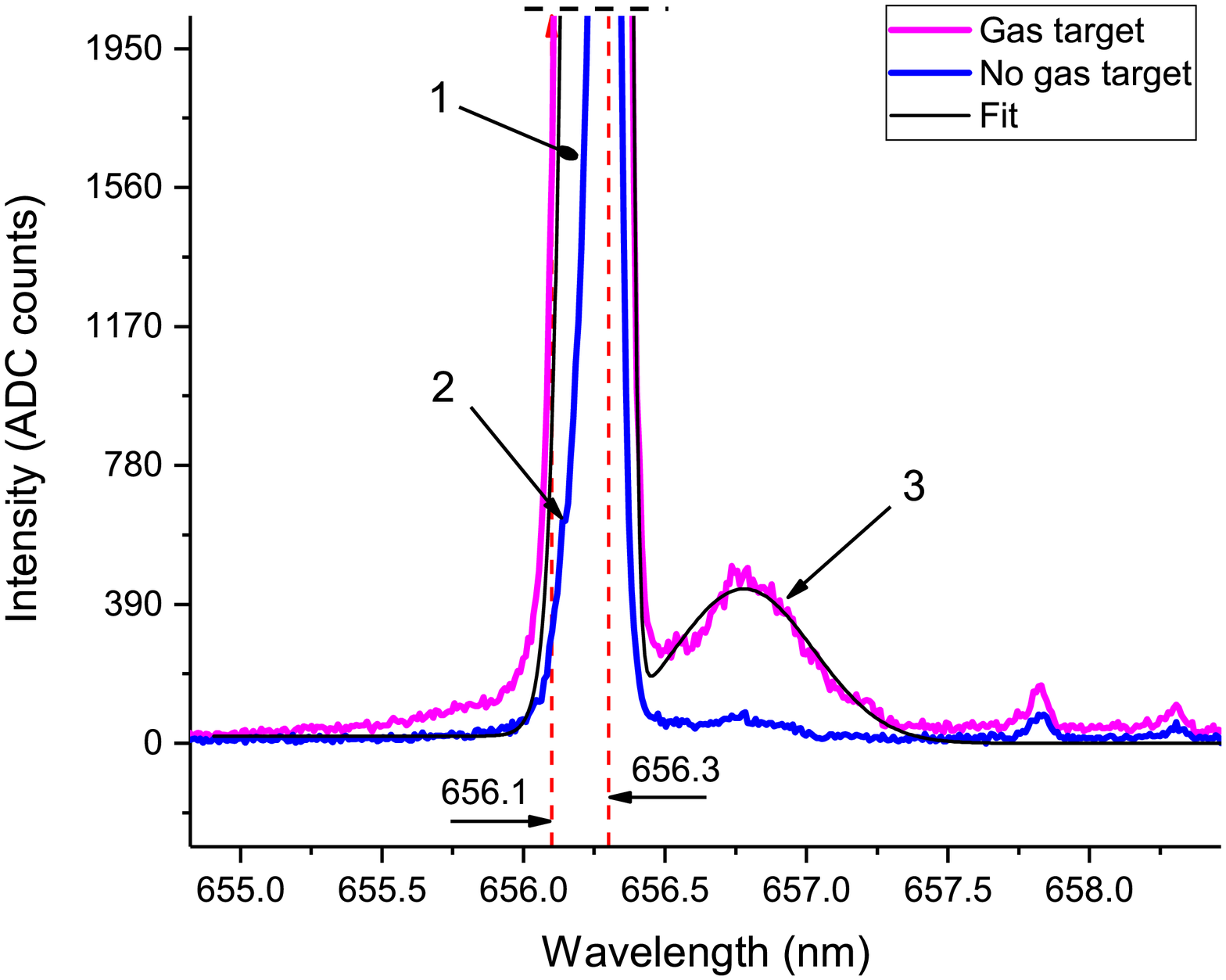}
\caption{\label{fig:Halpha_spectrum} Spectrum of H-$\alpha$ emission on the axis used for calculation of the mirror potential: magenta curve -- active CX frame acquired in the GDT shot 48994, blue curve -- passive frame acquired in the GDT shot 48993, black curve -- model fit of the active CX spectrum. Vertical dashed lines mark positions of unshifted D-$\alpha$ and H-$\alpha$. 1 -- bright cold-gas D-$\alpha$ and H-$\alpha$ lines (cut out),  2 -- D-$\alpha$ in the passive spectrum, 3 -- part of spectrum corresponding to the accelerated IDF. Acquisition timing parameters: $t=8~ms$, $\tau=0.5~ms$. }
\end{figure}
\paragraph{He-I spectral lines.}
The diagnostic scheme becomes more versatile with the additional helium puff in the central GDT part. Helium component is used as a tracer; the net He amount is small compared to the deuterium and hydrogen puff and it should be barely enough to create an observable optical signal. Atoms of He are stripped down to $He^{2+}$ with the characteristic time of $\sim 0.1 \div 0.5~ms$ which is smaller than the axial particle loss time $\tau_l \cong 1.5~ms$. One can expect the prevailing content of $He^{2+}$ over $He^+$ in the plasma flow in the expander, which is useful from the spectral analysis viewpoint due to the larger Doppler shift. One can not however presume the He ion population being in a local thermodynamic equilibrium with the bulk deuterium or deuterium-hydrogen plasma. Ions $He^{2+}$ are partially converted into $He^{*0}$ with subsequent emission of He-I lines. In this work, we observed the He-I lines 587.6~nm and 667.8~nm. The latter option was used in most shots because this He-I line fits the same working spectral range with H-$\alpha$. In this way, the measurements of the mirror potential, the central potential, the hydrogen temperature and the helium temperature were possible simultaneously.
The Figure~\ref{fig:full_spectrum} demonstrates the spectrum sample taken in the GDT shot 49311 with the charge exchange gas target switched on. The frame exposure was 0.5~ms recorded at $t=8~ms$ during the plasma heating phase. One may observe both the narrow cold-gas He line and the broader red-shifted emission responsible for accelerated $He^{2+}$ ions.
\begin{figure}[htbp]
\centering 
\includegraphics[width=.7\textwidth, trim=50 10 100 10,clip]{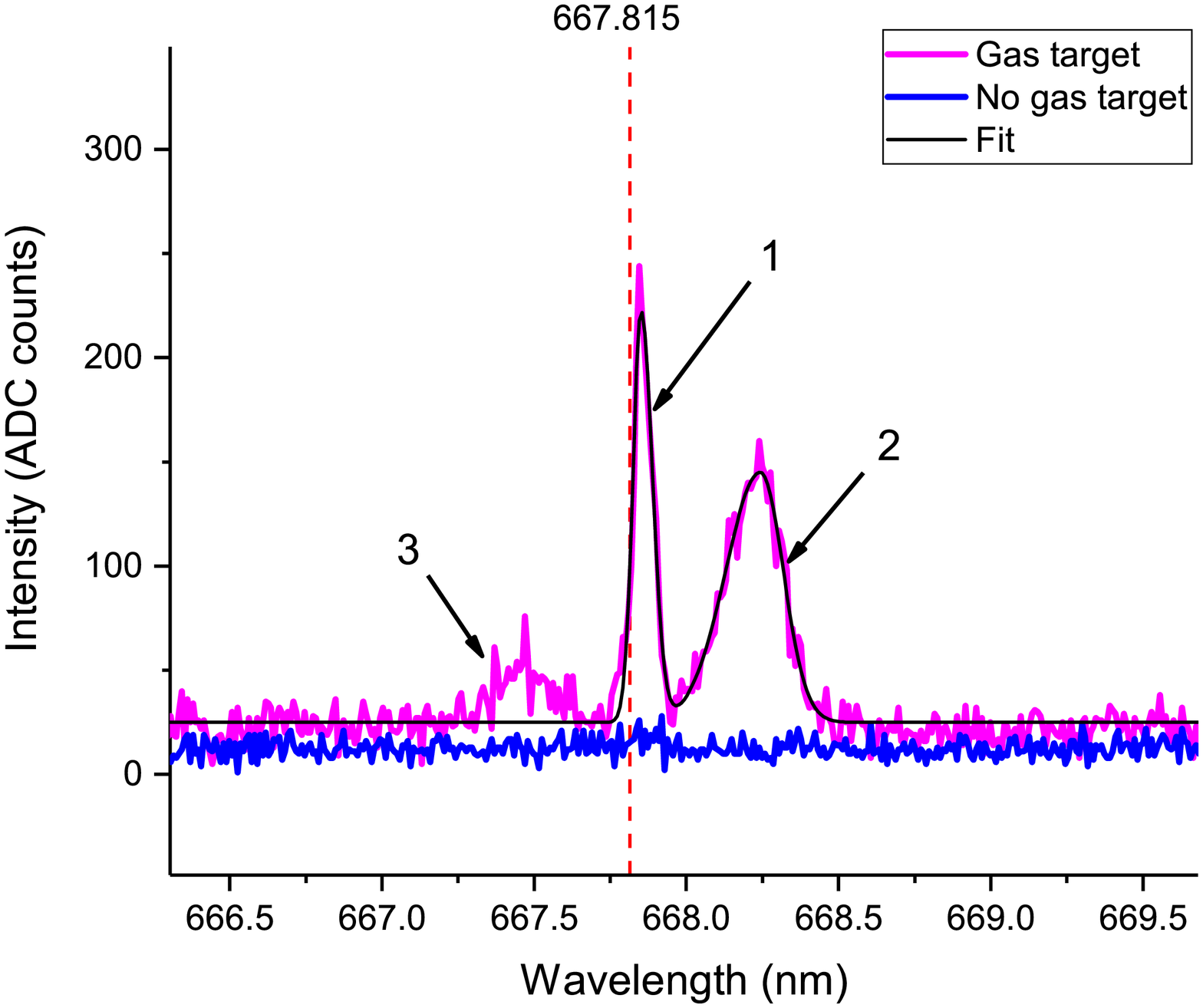}
\caption{\label{fig:He_spectrum} Spectrum of He-I emission 667.8~nm on the axis used for calculation of the central potential: magenta curve -- active CX frame acquired in the GDT shot 49707, blue curve -- passive frame acquired in the GDT shot 49705, black curve -- model fit of the active CX spectrum. The vertical dashed line marks the unshifted line position. 1 -- cold-gas He-I line,  2 -- active CX spectrum, 3 -- mirror reflection. Acquisition timing parameters: $t=8~ms$, $\tau=0.5~ms$. }
\end{figure}
\paragraph{Fitting CX spectra and measurement of potential and ion temperature.}
Figure~\ref{fig:Halpha_spectrum} shows the active CX H-$\alpha$ spectrum (magenta) measured on the axis in the shot 48994 and the background or passive spectrum from the shot 48993 (blue). The fit curve is also plotted (black). Note that the Doppler-shifted line is almost vanished in the passive sample. With this contrast ratio, we can neglect the passive contribution in the gas target enabled frame thus considering the recorded optical signal being the active CX emission. This ensures the spatial resolution considered above. In this particular series of shots, the hydrogen bulk plasma feed predominated. This lead to a relatively small background D-$\alpha$ emission, which is rendered in the passive spectrum as a smoothed dent (2) on the H-$\alpha$ wing, see Figure~\ref{fig:Halpha_spectrum}.
Physical data, namely the temperature and the potential, are obtained via fitting recorded spectra with the model function. The model we used, is a superposition of multiple bi-Gaussian lines each having its own set of parameters. Upon the fit convergence, line width and shift parameters are used for the calculation of the ion temperature and the potential drop via formulas \eqref{cxrs_doppler_width} and \eqref{cxrs_doppler_shift}. The mathematical processing codes are made using the nonlinear fit libraries provided in the Center Space NMath .NET software package \cite{nmath}. Error bars shown in all graphs that follow, reflect the accuracy of the spectrum fit with the model function.
Recorded spectra of He-I emission on the axis are plotted in the Figure~\ref{fig:He_spectrum}. Similar to the Figure~\ref{fig:Halpha_spectrum} with the H-$\alpha$ profile, both passive and active CX spectra are shown. There is no prominent Doppler-shifted signal above the noise level in the passive spectrum and likewise the H-$\alpha$ case, no background subtraction is needed for processing of the active CX spectrum. Easy to notice, that the cold-gas He-I line (1) also features the positive Doppler shift $\Delta\lambda_{cold}(He) \simeq 0.04~nm$ that is a reliable and reproducible effect. Such a shift is translated to the velocity $v_{cold}(He) \simeq 1.8\cdot 10^3~m/s$ which is consistent with collisions of cold He atoms in the expander with the accelerated plasma particles in the flow. The peak (3) we believe to be responsible for the mirror reflection of emitted light from the plasma dump surface. Due to some oversight, this surface is neither sand blasted nor darkened and the reflection would be remarkable indeed giving an approximately opposite Doppler shift as it is observable in the Figure~\ref{fig:He_spectrum}. We also admit a partial reflection of plasma flow particles from the dump surface.

\begin{figure}[htbp]
	\centering
	\begin{subfigure}{.45\textwidth}
		\includegraphics[width=\textwidth, trim=40 30 20 10,clip]{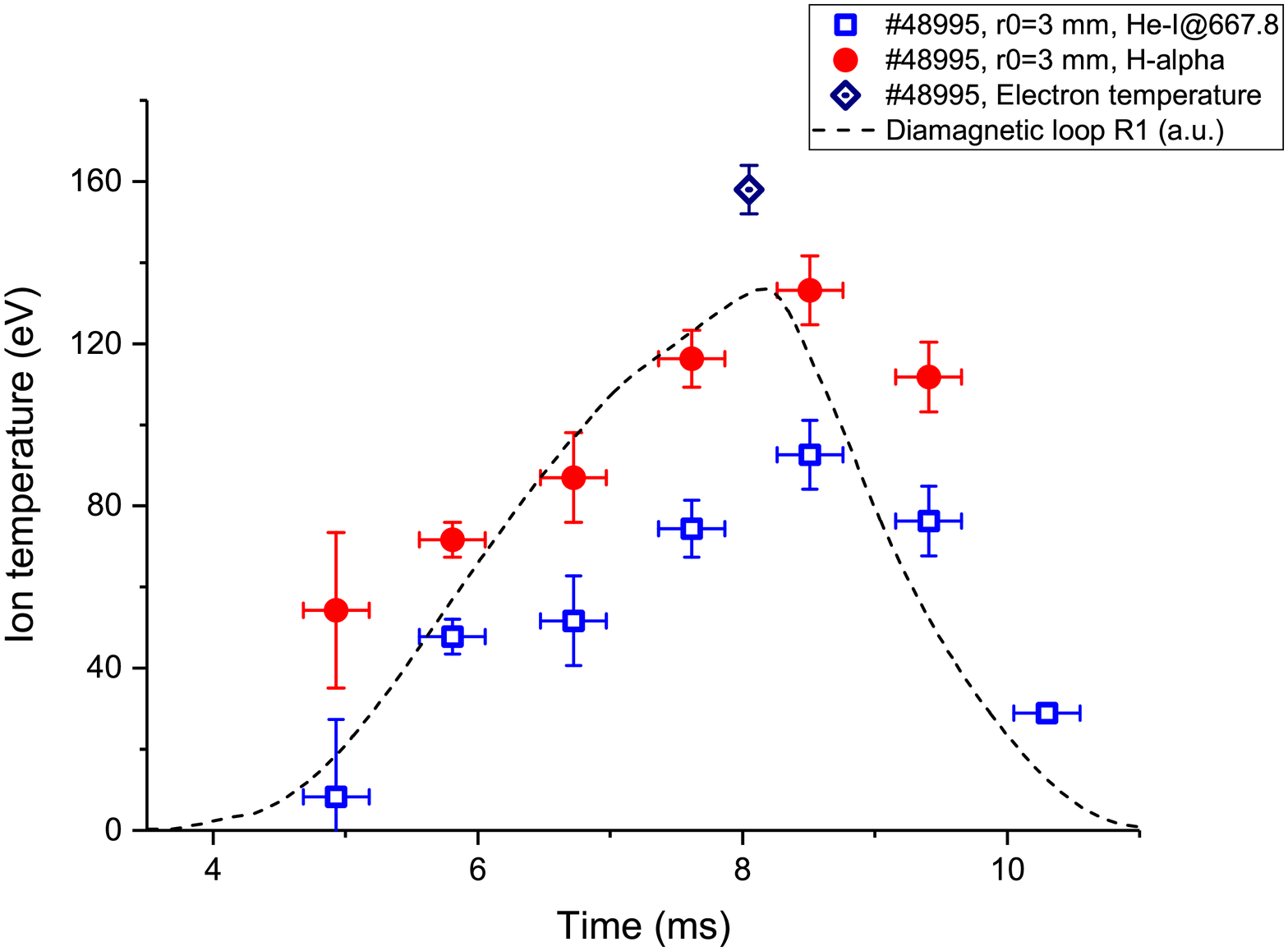}
		\caption{\label{fig:Ti} Ion temperature time evolution close to the axis. Horizontal error bars show the exposure duration of 0.5~ms. }
	\end{subfigure}
	\qquad
	\begin{subfigure}{.45\textwidth}
		\includegraphics[width=\textwidth, trim=50 30 20 10,clip]{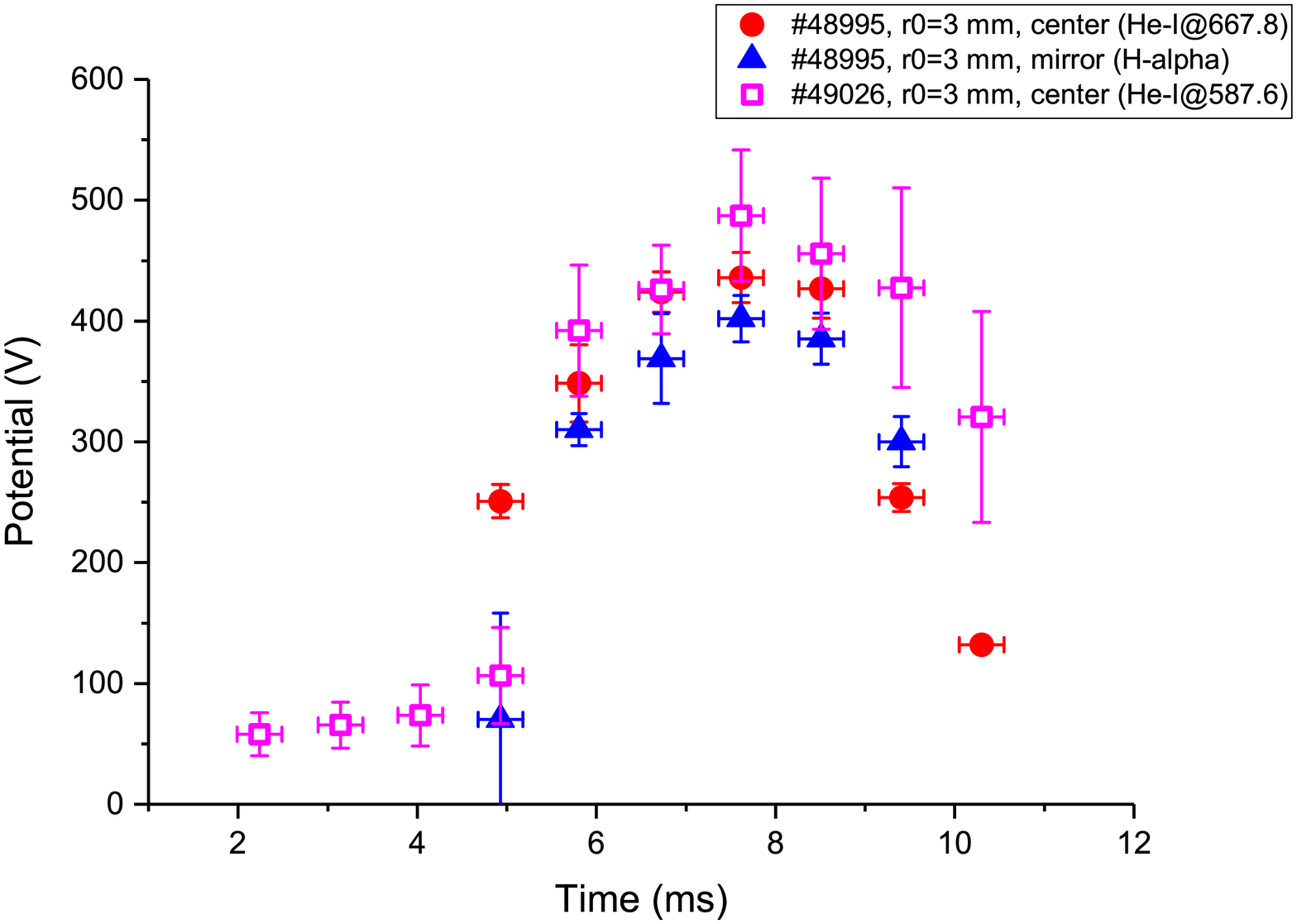}
		\caption{\label{fig:potential}  Time evolution of the plasma electrostatic potential in two locations (close to the axis). }
	\end{subfigure}
	\caption{Example of application of the CXRS diagnostic for the study of the ion velocity distribution function.}
\end{figure}

Using the spectrum mathematical analysis as explained above, time evolutions of the hydrogen ion temperature and $He^{2+}$ ion temperature are plotted in the Figure~\ref{fig:Ti}. The electron temperature is measured by the Thomson scattering in the GDT centre. The dashed curve provides the reference of diamagnetic signal showing the dynamics of the plasma energy content. The Figure~\ref{fig:potential} shows the time evolution of the plasma electrostatic potential measured via the H-$\alpha$ Doppler shift (blue filled triangles) and Doppler shifts of two He-I lines: 667.8~nm (red filled circles) and 587.6~nm (magenta open squares). The former two measurements are done simultaneously in the same shot, the latter one required tuning the spectrometer.

\section{Conclusion}
\label{conclusion}
The accuracy of considered CXRS measurements on He-I lines $\epsilon \approx 4 \div 8\%$ is primarily a function of the active signal intensity, because contribution of the background line emission along the LOS and the continuum light both are small. Mathematical calculations of the Doppler-shifted H-$\alpha$ are slightly complicated by the neighbouring cold-gas background line that may be approximately two orders of magnitude stronger. The exposure duration of 0.5~ms and the frame rate of 1.1~kHz allow for resolution of plasma parameters dynamics during the plasma startup, heating and sustainment in GDT. The spatial resolution of $\lesssim 5~mm$ determined by the gas cloud projection onto the middle plane, is sufficient for the study of spatial profiles of the ion velocity distribution function. In the present diagnostic design, there is a single line of sight which requires a series of shots for profile measurements. A comprehensive investigation of the axial transport of particles and energy in the gas dynamic trap with the intensive CXRS diagnostic involvement is under way. 

From the viewpoint of R\&D of new instrumentation, some valuable experience has been obtained as well. The basic diagnostic technique using the gas jet target is confirmed to be workable in a wide range of central cell plasma parameters and residual gas pressure in the expander vessel. It provides a solid ground for development of a more capable CXRS diagnostic version for the next-generation linear magnetic system for plasma confinement \cite{gdmt}, which construction is to be started within a couple of years. The improved optical system will have multiple observation points distributed across the plasma fan in the expander area. The charge exchange target should be a more collimated helium jet with a greater penetration depth, probably a supersonic gas jet like \cite{sgi-1, sgi-2}.

\acknowledgments
This work is supported by the Russian Science Foundation, project No.~18-72-10084 issued on 31.07.2018.


\begin{thebibliography}{99}
\bibitem{spitzer}
V.P.Pastukhov,
\emph{Nucl. Fusion}, {\bf 14} 3 (1974).

\bibitem{gdt-review-ppcf}
A. A. Ivanov and V. V. Prikhodko, 
\emph{Gas-dynamic trap: an overview of the
concept and experimental results}, 
\emph{Plasma Phys. Control. Fusion} {\bf 55} (2013) 063001.

\bibitem{nf-axconf-2020}
E.I. Soldatkina, V.V. Maximov, V.V. Prikhodko, V.Ya. Savkin, D.I. Skovorodin, D.V. Yakovlev and P.A. Bagryansky, 
\emph{Measurements of axial energy loss from magnetic mirror trap},
\emph{Nucl. Fusion} {\bf 60} (2020) 086009.

\bibitem{ecrh_prl}
Bagryansky P.A., Shalashov A., Gospodchikov E., Lizunov A., Maximov V., Prikhodko V., Soldatkina E., Solomakhin A. and Yakovlev D.,
\emph{Phys. Rev. Lett.}, {\bf 114} 205001 (2015).

\bibitem{sgi-1}
K. Schmid and L. Veisz,
\emph{Supersonic gas jets for laser-plasma experiments},
\emph{Rev. Sci. Instrum.}, {\bf 83}, 053304 (2012),
\url{http://dx.doi.org/10.1063/1.4719915}

\bibitem{sgi-2}
V. A. Soukhanovskii, H. W. Kugel, R. Kaita, R. Majeski, and A. L. Roquemore, 
\emph{Rev. Sci. Instrum.}, {\bf 75}, 4320 (2004),
\url{https://doi.org/10.1063/1.1787579}

\bibitem{he_ratio_nstx-u}
J. M. Mu\~noz Burgos, M. Agostini, et. al.,
\emph{Physics of Plasmas}, {\bf 23}, 053302, (2016),
\url{https://doi.org/10.1063/1.4948554}

\bibitem{he_ratio_rfx}
M. Agostini, P. Scarin, R. Cavazzana, A. Fassina, A. Alfier, and V. Cervaro,
\emph{Rev. Sci. Instrum.}, {\bf 81}, 10D715 (2010),
\url{http://dx.doi.org/10.1063/1.3478679}

\bibitem{he_ratio_aug}
M. Griener, E. Wolfrum, et. al.,
\emph{Rev. Sci. Instrum.}, {\bf 89}, 10D102 (2018),
\url{https://doi.org/10.1063/1.5034446}

\bibitem{he_ratio_hl2a}
B. D. Yuan, Y. Yu, et. al.,
\emph{Rev. Sci. Instrum.}, {\bf 91}, 073505 (2020),
\url{https://doi.org/10.1063/5.0005545}

\bibitem{gdt_expander_infulence}
Soldatkina E., Anikeev M., Bagryansky P., Korzhavina M.,
Maximov V., Savkin V., Yakovlev D., Yushmanov P. and
Dunaevsky A. 
\emph{Phys. Plasmas}  {\bf 24} (2017) 022505.

\bibitem{pylon-ccd}
\url{https://www.princetoninstruments.com/wp-content/uploads/2020/04/PyLoN_2k_datasheet.pdf}

\bibitem{nmath}
\url{https://www.centerspace.net/nmath}

\bibitem{gdmt}
A. Beklemishev,  A. Anikeev, et. al.,
\emph{Novosibirsk Project of Gas-Dynamic Multiple-Mirror Trap},
\emph{Fusion Science and Technology}, {\bf 63}, 1T (2013) 46-51.
\url{https://doi.org/10.13182/FST13-A16872}

\end{thebibliography}
\end{document}